\documentclass[12pt]{article}

\usepackage{amsmath, amssymb, fullpage}
\usepackage{natbib}

\begin{document}
\begin{center}
\textbf{\large{Discussion on Using Stacking to Average Bayesian Predictive Distributions by Yao et al.}}\\ \vspace{10pt}
\textbf{William Weimin Yoo}\\
\textit{Leiden University}
\end{center}

\begin{abstract}
I begin by summarizing key ideas of the paper under discussion. Then I will talk about a graphical modeling perspective, posterior contraction rates and alternative methods of aggregation. Moreover, I will also discuss possible applications of the stacking method to other problems, in particular, aggregating (sub)posterior distributions in distributed computing.
\end{abstract}

\noindent\textbf{Keywords:} Stacking, Posterior contraction rate, Prediction and credible sets, Fourier transform, Distributed computing\\ \vspace{10pt}

In this paper, \citet{gelman} consider the problems of model selection and aggregation of different candidate models for inference. Inspired by the stacking of means method proposed in the frequentist literature, the authors generalize this idea by developing a procedure to stack Bayesian predictive distributions. Given a list of candidate models with their corresponding predictive distributions, the goal here is to find a linear combination of these distributions such that it is as close as possible to the true date generating distribution, under some score criterion. To find the linear combination (model) weights, they replace the full predictive distributions with their leave-one-out (LOO) versions in the objective function, and proceed to solve this convex optimization problem. The authors then propose a further approximation to LOO computation by importance sampling, with the importance weights obtained by fitting a generalized Pareto distribution. To showcase the benefits of the newly proposed stacking method, the authors conducted extensive simulation studies and data analyses, by comparing with other techniques in the literature such as BMA (Bayesian Model Averaging), BMA with LOO (Psedo-BMA), BMA with LOO and Bayesian Bootstrap, and others.

We can take a graphical modeling perspective on LOO. For example, replacing marginal likelihoods $p(y|M_k)$ with $\prod_{i=1}^np(y_i|y_j:j\neq i, M_k)$ is akin to simplifying a complete (fully connected) graph linking observations to one where the Markov assumption holds, i.e., the node corresponding to $y_i$ is independent conditioned on its neighbors $\{y_j:j\neq i\}$. In the proposed stacking method, the full predictive distribution $p(\widetilde{y}|y)$ is approximated by the LOO $p(y_i|y_j:j\neq i)$, and further approximation is needed because the LOO is typically expensive to compute. However if there are some structures in the data, such as clusters of data points/nodes, then one can take advantage of this by conditioning on a smaller cluster $\mathcal{B}$ of nodes around $y_i$ and compute instead $p(y_i|y_j:j\neq i,j\in\mathcal{B})$. This would then further speed up computations as one can fit models using only local data points.

Another point I would like to make is that the superb performance of the stacking method warrants further theoretical investigations. Figure 2(c) in the simulations shows that the proposed method is robust to incorrect/bad models, in the sense that its performance stays unchanged even if more incorrect models are added to the list. It would be nice if we will also have some theoretical guarantees that the stacking method will concentrate on the correct ($\mathcal{M}$-closed) or the best models ($\mathcal{M}$-complete) in the model list. In addition, Figure 9 shows that this method is able to ``borrow strength" across different models, by using some aspects of a model to enhance performance of a different model. Therefore aggregation by stacking adds value by bringing the best out of each individual model component, and it would be interesting to characterize through theory what this added value is. This then invites us to reflect on how the quality of individual model component affects the final stacked distribution. For example, given posterior contraction rates for the different posterior models, what would be the aggregated rate for the stacked posterior? Also, how does prediction/credible sets constructed using the stacked posterior compare with those constructed from each model component, will it be bigger, smaller or something in between?

As most existing methods including the newly proposed stacking method use some form of linear combinations, it would be interesting to find other ways of aggregation. As pointed out by the authors, linear combinations of predictive distributions will not be able to reproduce truths that are generated through some other means, e.g., convolutions. To apply stacking in the convolutional case, I think one way is to do everything in the Fourier domain, by stacking log Fourier transforms (i.e., log characteristic functions) of the predictive posterior densities, exponentiate and then apply inverse Fourier transform to approximate the truth generated through convolutions.

I think another possible area of application for the stacking method is in distributed computing. In this modern big data era, data has grown to such size that it is sometimes infeasible or impossible to analyze them using standard Markov Chain Monte Carlo (MCMC) algorithms on a single machine. Hence this gave rise to the divide and conquer strategy where data is first divided into batches and (sub)posterior distribution is computed for each batch. The final posterior for inference is then obtained by aggregating these (sub)posteriors. To this end, I think the present stacking method can be deployed after some modifications, with potential to yield superior performance when compared with existing weighted average-type approaches.

I find the paper to be very interesting and the stacking method proposed is a key contribution to the Bayesian model averaging/selection literature. It is shown to be superior than the golden standard, i.e., BMA and its finite sample performance is tested comprehensively though a series of numerical experiments and real data analyses. I think this is a very promising research direction, and any future contributions are very welcomed.

\bibliographystyle{apa}
\bibliography{predictive}

\begin{thebibliography}{}

\bibitem[\protect\astroncite{Yao et~al.}{2018}]{gelman}
Yao, Y., Vehtari, A., Simpson, D., and Gelman, A. (2018).
\newblock Using stacking to average {Bayesian} predictive distributions.
\newblock {\em Bayesian Anal.}, pages 1--28.
\newblock Advance publication.

\end{thebibliography}
\end{document}